\documentclass{aa}  

\usepackage{graphicx}
\usepackage{txfonts}
\usepackage{textcomp}
\usepackage{amsmath}
\usepackage{xspace}   % for \xspace
\usepackage{xcolor} 
\newcommand{\software}[1]{\texttt{#1}}
\newcommand{\unit}[1]{\ensuremath{\mathrm{\,#1}}\xspace}
\newcommand{\kms}{\unit{km\,s^{-1}}}
\newcommand{\masyr}{\unit{mas\,yr^{-1}}}
\newcommand{\kpc}{\unit{kpc}\xspace}
\newcommand{\gaia}{{\it Gaia}\xspace}

\newcommand{\teff}{\ensuremath{T_{\mathrm{eff}}}}

\newcommand{\feh}{\unit{[Fe/H]}}
\newcommand{\sag}{Sgr~A$^\ast$\xspace}
\newcommand{\agama}{\texttt{AGAMA}\xspace}
\newcommand{\fcomp}{\ensuremath{10^{-4}}}

\newcommand{\candidate}{DESI-312\xspace} %to run a race repetitively and constantly. 

\newcommand{\ratiogc}{\texttt{ratio\_gc}\xspace}

\usepackage{natbib}
\usepackage{hyperref}

\begin{document}

\title{Discovery of Galactic center ejected star in DESI DR1}

   %\subtitle{Southern Hemisphere counterpart}

   \author{Manuel Cavieres\inst{1}\fnmsep\thanks{Corresponding author: Manuel Cavieres\\
          \email{cavierescarrera@strw.leidenuniv.nl}}
          \and
          Sergey E.~Koposov\inst{2,3}
          \and 
          Elena Maria Rossi\inst{1}
          \and 
          Zephyr Penoyre\inst{1}
          \and
          Sill Verberne\inst{1}
          }

      \institute{Leiden Observatory, Leiden University,
              P.O. Box 9513, 2300 RA Leiden, the Netherlands
         \and
             Institute for Astronomy, University of Edinburgh, Royal Observatory, Blackford Hill, Edinburgh EH9 3HJ, UK
         \and
             Institute of Astronomy, University of Cambridge, Madingley Road, Cambridge CB3 0HA, UK
             }
   \date{Received September 15, 1996; accepted March 16, 1997}

% \abstract{}{}{}{}{} 
% 5 {} token are mandatory
 
  \abstract
   {Hypervelocity stars (HVSs) are stars ejected from the Galactic Centre (GC) through tidal interactions with the central supermassive black hole. Formed in the immediate vicinity of Sgr~A$^\ast$, these stars are accelerated to velocities high enough to escape the GC and be observable in the Galactic halo. Using spectroscopy from the Dark Energy Spectroscopic Instrument (DESI) and astrometry from \textit{Gaia}, we conducted a six-dimensional search for HVSs and identified a compelling candidate, hereafter \candidate, whose bound trajectory can be confidently traced back to the central 2\kpc of our galaxy. The star resides in the inner halo and exhibits supersolar metallicity (\feh $= 0.27\pm 0.09$), distinct from other known stellar populations with radial orbits. Its inferred GC ejection velocity of $698^{+35}_{-27}$ is consistent with a Hills mechanism ejection, supporting an origin in the innermost regions of the Milky Way. We considered alternative origins for the star, including disk ejections from young clusters and globular clusters, but these scenarios fail to explain both its orbit and metallicity. Unlike previously identified A- and B-type HVSs, \candidate\ is a $\sim 1\,M_{\odot}$ star on the main sequence or early subgiant branch, thus enabling a detailed chemical analysis of its atmosphere and offering a rare window -- unobscured by dust and crowding -- into the composition of the central regions of the Galaxy.}


   \keywords{ Stars: Black holes, kinematics and dynamics; Galaxy: nucleus
               }

   \maketitle
%
%-------------------------------------------------------------------

\section{Introduction}

  %Hypervelocity stars (HVSs) are stars ejected from the Galactic Centre (GC) through binary interactions with the massive black hole (MBH) in the GC, Sagittarius A* (Sgr A*). In this process, known as the Hills mechanism \citep{hills}, one star is captured by Sgr A*, while the other is ejected at velocities of up to several thousand \kms. T

Hypervelocity stars (HVSs) are stars ejected from the vicinity of Sgr~A$^\ast$ in our Galactic Centre (GC); when the ejection velocity is of the order of $\sim 1000$ \kms, they can reach the halo of our Galaxy. The most promising mechanism for producing such high velocities is the Hills mechanism \citep{hills}, in which a stellar binary is tidally disrupted by an interaction with a massive black hole, yielding one captured component and one ejected star with velocities up to thousands of \kms. A similar scenario ejects single stars following the interaction with a binary massive black hole; this assumes the existence of an intermediate black hole companion orbiting around Sgr A* \citep[eg.,][]{bbh_ejection, 2006ApJ...653.1203L, 2006ApJ...651..392S, 2009MNRAS.392L..31S, 2019ApJ...878...17R}.

Alternative channels include dynamical encounters in dense stellar systems \citep{1967BOTT....4...86P, 1990AJ.....99..608L, 2009MNRAS.396..570G} and supernova explosions in binary systems \citep{1961BAN....15..265B, 2000ApJ...544..437P}. Both processes have been observed in the Galaxy  \citep{2001A&A...365...49H, 2025arXiv250910387Z}, with typical ejection velocities exceeding $\sim 40$ \kms \citep{1961BAN....15..265B, 2012ApJ...751..133P}, defining the category of runaway stars. Simulations predict that most runaway stars from dynamical encounters have ejection velocities $\leq 200$ \kms \citep{2012ApJ...751..133P}, while faster unbound ejections (often termed hyper-runaway stars) are extremely rare, with a rate of $8 \times 10^{-7} \text{yr}^{-1}$ \citep{2012ApJ...751..133P, review_brown}. Therefore, the primary mechanism for accelerating stars to velocities exceeding $200$ \kms is an interaction with Sagittarius A*.

These ejected stars experience the entire Galactic potential, from the GC to the outskirts of the Milky Way (MW) halo, making them prime tracers of the Galactic potential \citep{2005ApJ...634..344G, 2007MNRAS.379.1293Y, review_brown, 2022A&A...663A..72G}, and even possible probes of modified theories of gravity \citep{2022A&A...657A.115C}. 

Importantly, HVSs provide a unique opportunity to study the properties of the GC, including the star formation history, merger history, and chemical composition. Due to high dust extinction, direct chemical abundance measurements within the GC are extremely challenging, particularly in the central pc \citep{nscchemical}. Because HVSs originate from these inner regions but travel beyond the dusty environment of the GC, they provide a powerful means to indirectly study the chemical makeup of their progenitor populations. Since the HVS ejection mechanism is intrinsically linked to the capture of objects around massive black holes, depending on initial conditions \citep{bianca_25}, the ejection rate of HVSs is crucial for understanding stars in the vicinity of SgrA*, such as the S-star cluster \citep{sill_s_stars}; Additionally,   approximately 20\% of captured stars may result in a tidal disruption event (TDE) \citep{sill_s_stars}; if the captured object is instead a stellar mass black hole, it can be losing orbital energy via emission of gravitational waves and be detected by LISA \citep{LISA} as an extreme mass ratio inspirals (EMRI); Consequently, the HVS ejection rate plays a key role in constraining both the EMRI and TDE rates.

Many HVS searches have been conducted, primarily using \gaia data \citep[e.g.,][]{tomassodr2,6dtomasso_1,6dtomasso,alonso_luna_2024,Sill, hattori25}, or via the photometric identification of faint blue stars \citep{hvs_survey}. These efforts have enabled the identification of some of the most promising candidates to date, such as unbound HVS1 \citep{brown2005}, which, however, lacks sufficient astrometric precision to be definitively associated with the GC. Nevertheless, HVSs are rare: only about a dozen promising candidates have been identified, and a single unambiguous HVS has been detected to date---S5-HVS1 \citep{s5hvs1}.

Relying only on \gaia data limits the HVS search as the subset of stars in \gaia with measured radial velocities represents only a small fraction of the full \gaia catalogue: this sample is restricted to the brightest ($G_{RVS}<16$ mag) objects, and lacks chemical information that would allow one to separate the HVS candidate from halo populations in similar orbits. In this context, the advent of multi-object spectroscopic surveys such as 4MOST \citep{2019NatAs...3..574D}, WEAVE \citep{2012SPIE.8446E..0PD}, and, most importantly for this work, DESI \citep{desidr1}, represents the next step in HVS searches, increasing sensitivity by targeting fainter objects and providing radial velocities and stellar parameters for millions of stars across the sky.

The first data release of the DESI survey \citep{desidr1} includes spectra for more than 18 million unique targets, mainly aimed at mapping the evolving three-dimensional structure of the Universe up to $z = 4$. In addition, through the Bright programme of the main survey and the backup programme, DESI has observed more than 4 million stars. Measurements of stellar parameters, abundances, and radial velocities for these stars are provided in the stellar value-added catalogue \citep{DESIMW}. Furthermore, \citet{desi_dist} constructed the value-added distance catalogue \textit{Specdis}, which provides distances for objects in the stellar catalogue using a feed-forward neural network trained directly on DESI spectra. The combined catalogue includes radial velocities from the DESI stellar catalogue, spectroscopic distances from the \textit{Specdis} catalogue, and proper motions and sky positions from the third data release of the \gaia mission \citep{gaiadr3}. This combined dataset provides full six-dimensional phase-space information for approximately 4 million stars

Additionally, the determination of metallicities from DESI spectra is crucial for an HVS search, as metal-poor halo stars are the principal contaminants. Combining chemical abundances with kinematics to constrain stellar birth locations via chemical tagging increases the reach of HVS searches, as it opens a window to studying bound HVS—i.e., stars ejected from the GC whose total velocities are insufficient to escape the gravitational field of the Galaxy—which are expected to dominate over the population of unbound HVS \citep{review_brown,2018MNRAS.476.4697M, evans_lmc}. An example of this type of object is WINERED-HVS1 \citep{hattori25}, a bound red giant with \feh $\sim -0.15$ whose trajectory can be traced back to the GC and implies a modest ejection velocity of 500 \kms. Its abundance pattern in [$\alpha$/Fe], [Mg/Fe], [Si/Mg], and [Ca/Mg] associates it with the nuclear star cluster (NSC); however, given its metallicity and low velocity, it is also considered an ambiguous HVS as the metallicity makes it likely to be a halo star.

In this work, we performed a 6D search for HVSs ejected from the GC, including those bound. This is done by combining radial velocities and metallicities from the DESI Data Release 1 stellar catalogue \citep{DESIMW}, \gaia DR3 \citep{gaiadr3} proper motions and sky positions, along with spectroscopic distances from the SpecDis \citep{desi_dist} value-added catalogue. 

This paper is organized as follows. In Section \ref{sec:data}, we describe the data used in our HVS search, while in Section \ref{sec:methods}, we detail the methodology used to identify HVS candidates, which can be extended to any 6D dataset.  We describe our discovery of a strong HVS candidate in Section \ref{sec:good_mw_candidate}. We discuss its possible origins along with a comparison with previous studies in Section \ref{sec:discussions}. Finally, we draw our conclusions in Section \ref{sec:conclusions}.
%--------------------------------------------------------------------

\section{Data}\label{sec:data}

%We have previously discussed the method for selecting hypervelocity stars ejected from the Galactic center. Here, we describe the data from DESI and \gaia used in this study, as well as the quality selections applied.
In this Section, we describe the observational data from DESI and \gaia used to search for HVSs, as well as the quality selections applied to it. The simultaneous use of these two surveys allow us to gather full phase- space information as well as metallicity measurements for each examined star.

We utilized data from the first data release of DESI, particularly the DESI stellar catalogue (MWS) \citep{DESIMW}, which contains spectra and stellar properties for more than $\sim 4$ million stars. Radial velocities and \feh\ measurements were taken from the RVS pipeline \citep{rvspecfit1, rvspecfit2}, and combined with distances from the spectrophotometric \textit{Specdis} catalogue \citep{desi_dist}. The \textit{Specdis} catalogue provides distances for MWS stars using a feed-forward neural network trained directly on DESI spectra. For the \feh\ measurements, we adopted the calibration described in \citet{DESIMW} to account for biases in the RVS pipeline.To obtain full 6D phase-space information, we combined DESI data with \gaia\ astrometry using the crossmatch included in the DESI stellar catalogue. 
Additionally, to ensure that we are working with reliable measurements, we imposed quality selections, which are summarized in Table~\ref{tab:quality}. Briefly, we select sources with $\mathrm{RUWE} < 1.4$ to exclude objects with unreliable astrometry. Additionally, we ensure that objects are classified as stars by the \texttt{redrock} classifier \citep{redrock} by requiring $\mathrm{RR\_SPECTYPE} = '\mathrm{STAR}'$, and we remove stars with potentially problematic spectral fits by requiring $\texttt{RVS\_WARN} = 0$.

\begin{table}[h]
\centering
\caption{Quality selections.} \label{tab:quality}
\begin{tabular}{lc}
    \hline
    Selection Criteria \\
    \hline
    $\texttt{RR\_SPECTYPE} = \texttt{'STAR'}$, \\
    $\texttt{RUWE} < 1.4$, \\
    $\texttt{RVS\_WARN} = 0$, \\
    %$\texttt{DIST}/\texttt{DISTERR} > 5$ \\
    \hline
\end{tabular}
\end{table}

\section{Methods}\label{sec:methods}
In this Section, we describe the method for selecting hypervelocity stars ejected from the GC using the six-dimensional phase space measurements from DESI DR1 and \gaia, presented in the previous Section.

\subsection{Galactic Centre origin}\label{subsec:mw_origin}
To identify hypervelocity star candidates that could originate from the GC of the MW, we first transform their sky position, distance, proper motion, and radial velocity into the Galactocentric reference frame using the Python package \software{Astropy \citep{astropy:2013, astropy:2018, astropy:2022}}. This transformation assumes a distance to Sgr~A* of 8.122~\kpc, a solar three-dimensional velocity of (12.9, 245.6, 7.78)~km\,s$^{-1}$, and a solar height above the Galactic plane of 20.8~pc. We then select objects with galactocentric radial velocity $|V_{rf}| > 300$ km\,s$^{-1}$, which removes most objects associated with the MW disc, while retaining relatively fast but still bound galactic centre ejecta. This selection yields 54.475 stars from DESI.

In order to investigate these stars' origin, we used the Python package \software{AGAMA \citep{agama}} to perform backward orbit integration. We adopted the Price-Whelan 2017 potential, which corresponds to a porting of Gala \citep{gala} MilkyWayPotential2022 to \texttt{AGAMA}. This is a mass model for the MW composed of a spherical nucleus, a bulge, a sum of three components of Miyamoto-Nagai discs \citep{MN_disk} to represent an exponential disc, and a spherical NFW dark matter halo \citep{NFW}. This model is tuned to reproduce both the MW rotation curve from \citet{2019ApJ...871..120E} and the vertical structure from the phase spiral in the solar neighbourhood of \citet{2023ApJ...955...74D}.

For each object, we generated 1000 Monte Carlo samples obtained by Gaussian sampling over uncertainties in radial velocity, proper motion, and distance. Each sample is integrated backward in time over 0.5~Gyr using adaptive time steps in the \texttt{AGAMA} framework. During integration, the positions where the orbit crosses the Galactic plane are recorded. We then compute the GC crossing ratio, denoted as \ratiogc, defined as the fraction of samples that cross the plane within the circle of 1~\kpc of radius around the GC. Here, a star with a \ratiogc of 0.5 has a 50\% chance of having its first plane crossing within 1 \kpc from the GC.

We treat any star with \texttt{ratio\_gc} $\geq 0.5$ as a candidate GC ejecta. Unlike most previous HVS searches \citep[e.g.,][]{tomassodr2,6dtomasso_1,6dtomasso,alonso_luna_2024,Sill}, we do not exclude bound objects (see also \citet{hattori25}). However, a purely kinematic selection that admits bound objects is vulnerable to contamination by accreted populations, and potentially eccentric in-situ stars if they exist \citep{1962ApJ...136..748E}. In particular, stars associated with the Gaia–Sausage–Enceladus merger \citep[GSE;][]{gse_helmi,gse_vasily} often occupy the same phase-space region as HVS candidates, as they are typically on highly eccentric, radial, and often retrograde orbits. Fortunately, GSE stars have a well defined metallicity distribution that peaks at \feh $\sim -1.6$ and quenched at \feh $\sim -0.5$ \citep{2024A&A...691A.333E}. Therefore, to mitigate contamination from GSE stars, we restricted our sample to candidates with supersolar metallicity (\feh $> 0$).

\begin{figure}  
    \centering
    \includegraphics[width=\linewidth]{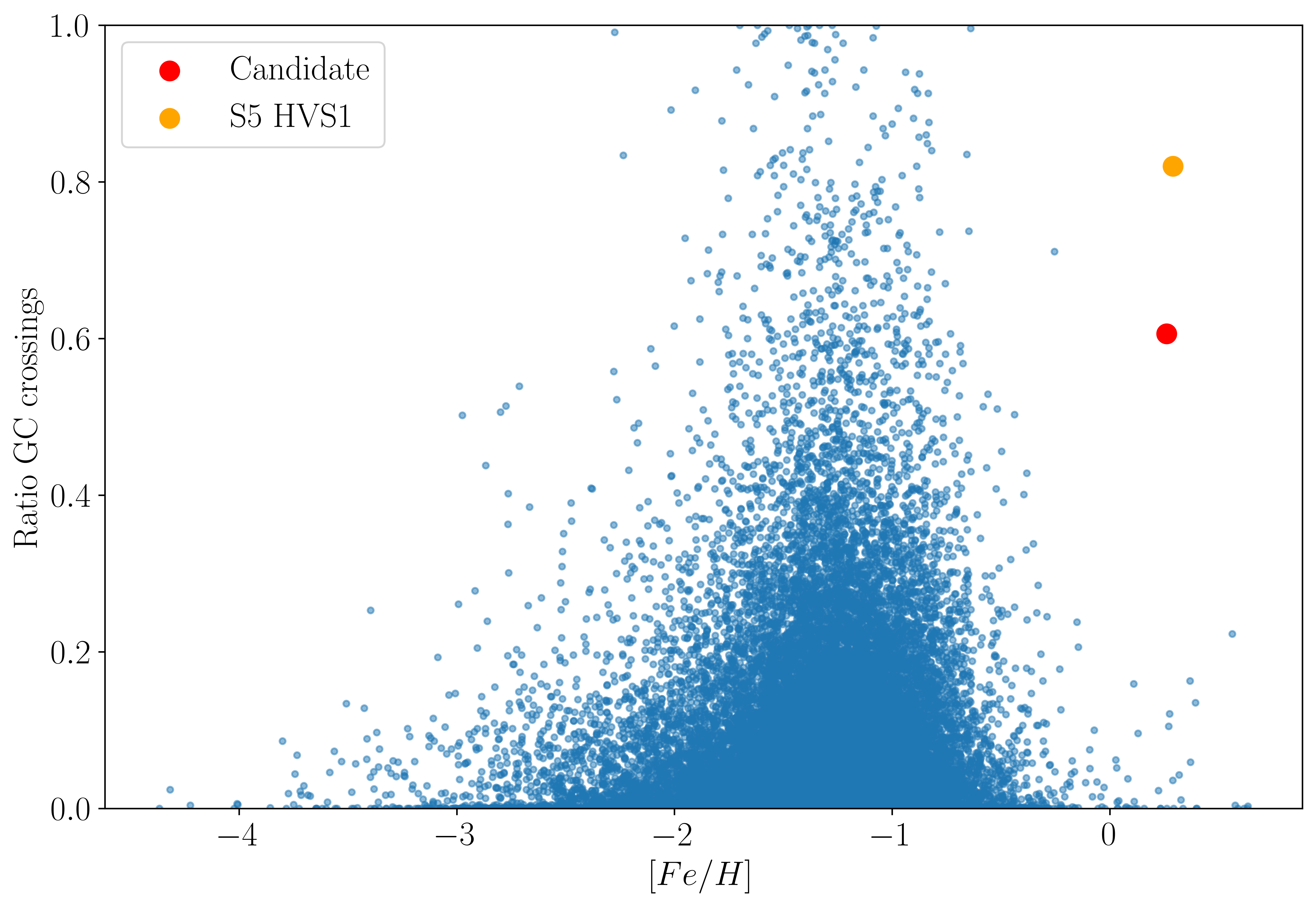}
    \caption{Fraction of realizations (out of 1,000 total) whose first Galactic-plane crossing falls within a 1 \kpc-radius circle around the GC, shown as a function of \feh for DESI candidates. S5-HVS1 is overplotted in orange for reference \citep{s5hvs1}. The only candidate found, star \candidate, is shown in red.
 }
    \label{fig:feh_gc_ratio}
\end{figure}

\section{Results} \label{sec:results}

By applying the methodology described in Section \ref{sec:methods} to the data set described in Section \ref{sec:data}, we obtained the distribution of \ratiogc shown in Figure \ref{fig:feh_gc_ratio}. Here, the requirement of supersolar metallicity ($\feh \geq 0 $) removes all but one object from the purely kinematic selection, shown in red in Figure \ref{fig:feh_gc_ratio}. Therefore, our analysis yields a single candidate, star SOURCE ID 4226880901740499200, to which we will refer as \candidate, and will be described in Section \ref{sec:good_mw_candidate}. 

Additionally, Figure~\ref{fig:feh_gc_ratio} shows that the majority of high \ratiogc objects have metallicities consistent with a GSE origin. For reference, the unambiguous HVS, S5-HVS1, is also shown, which has a metallicity of \feh $0.29 \pm 0.08$ dex and a \ratiogc obtained by applying our search's methodology to the parameters of S5-HVS1 presented in \citep{s5hvs1}.

\subsection{HVS candidate \candidate} \label{sec:good_mw_candidate}
Having described the HVS selection methodology and the data on which it is applied, in this Section, we describe the properties of the only HVS candidate found, \candidate, as determined from DESI DR1 data, \gaia astrometry, and available photometric data. The summary of the \gaia and DESI measurements is presented in Table \ref{tab:desi_hvs1}, and the photometric measurements used are presented in Table \ref{tab:photometric_measurements}. 

\subsubsection{Spectroscopy}
\begin{figure*}
    \centering
    \includegraphics[width=0.9\linewidth]{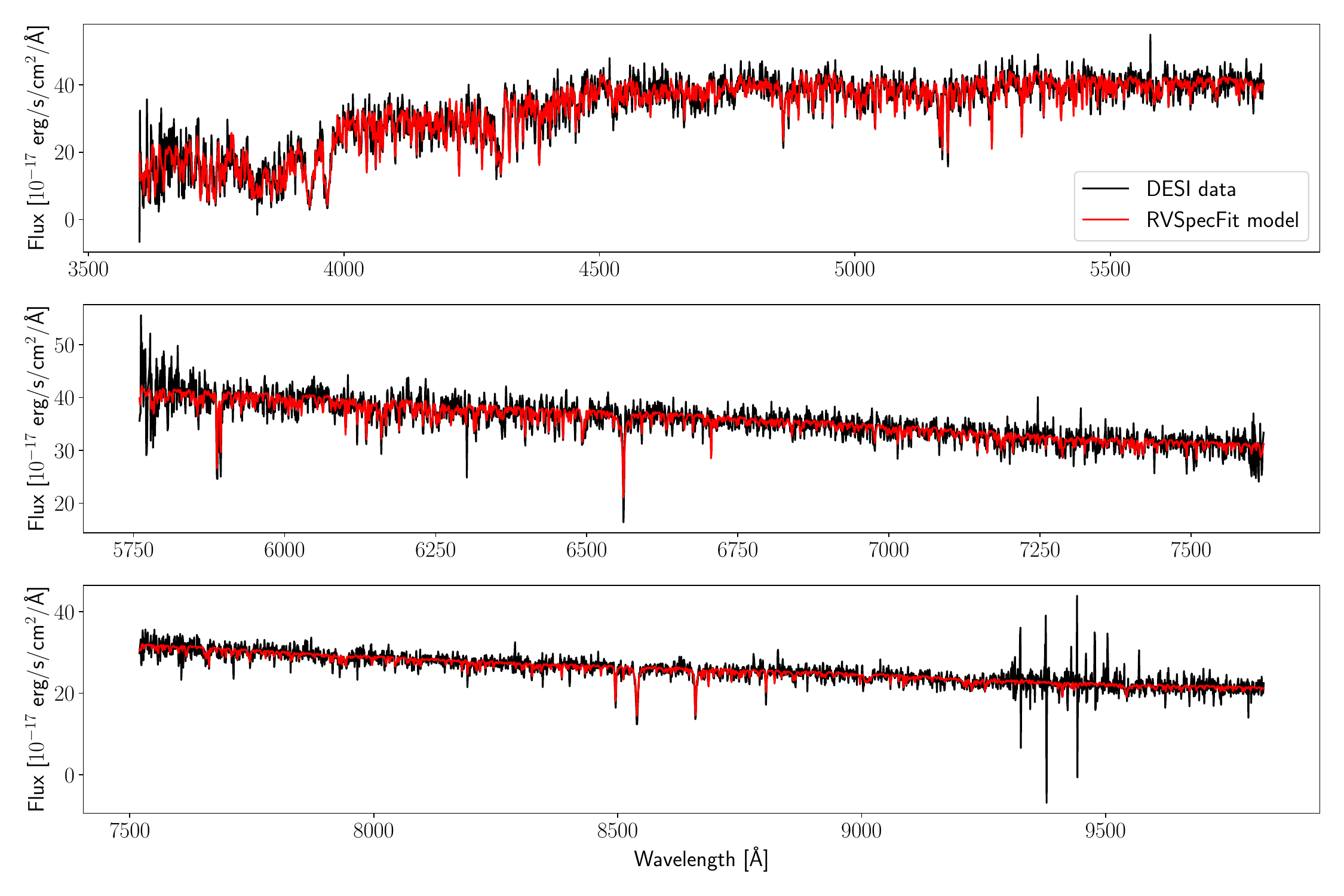}
    \caption{Comparison between the best-fitting model spectrum (red) and the observed DESI spectra (black) for \candidate, both provided as part of the first data release of the DESI stellar catalogue \citep{DESIMW}. From top to bottom, the panels show the red, blue, and z arms of the spectra. The close agreement between model and data gives no suggestion of a problematic fit or unreliable atmospheric parameters.}
    \label{fig:spectra_source}
\end{figure*}

This source was observed as part of the bright program during the main DESI survey. Being a relatively bright source with \gaia G $\sim$ 17, the observed spectra were obtained with a signal-to-noise ratio of $\sim$ 30.5 in the R and Z arms, and $\sim 19$  in the blue arm. The spectra, along with the best fit PHOENIX \citep{phoenix} model from rvspecfit \citep[][]{rvspecfit1, rvspecfit2}, released with the DESI stellar catalogue, are shown in Figure \ref{fig:spectra_source}.

\begin{table}
    \caption{Measured parameters of \candidate from \gaia and DESI DR1. Parallax corresponds to the zero-point corrected value. Reported \feh corresponds to the corrected metallicity \citep{DESIMW}.}
    \label{tab:desi_hvs1}
    \centering
    \begin{tabular}{c|c|c}
    \hline
    Parameter & Value & unit \\
    \hline
    \gaia RA & $312.7904997$ & deg \\
    \gaia Dec &$-1.00806508$ & deg \\
    %\gaia DR3 {\tt source\_id} & $4226880901740499200$ &  \\
    \gaia $\mu_\alpha \cos \delta$ & $-12.6 \pm 0.0864$ & \masyr \\
    \gaia $\mu_\delta$ & $-13.8 \pm 0.069$ & \masyr \\
    \gaia Parallax & $0.231 \pm 0.082$ & mas \\
    \gaia RUWE & $1.03$ &  \\
    ${\it E(B-V)}_{\rm SFD}$ & $0.0859$ &  \\
    \gaia G & $17.1246$ & mag \\
    \hline
    HRV & $-172 \pm 0.62$ & \kms \\
    \teff & $5579.3 \pm 29.1$ & K \\
    $\log g$ & $4.21 \pm 0.0651$ & dex \\
    \feh & $0.263 \pm 0.027$ & dex \\
    $v\sin{i}$ & 0.01 & \kms \\ 
    \hline
    $D_{hel}$ & $4.78 \pm 0.83$ &  \kpc \\

    \hline
    % here add the 3d velocity of the star
    %$X_{gal}$ & $-5.20 \pm 0.50$ & \kpc
    $V_{gcr}$ & $ 313.94 \pm 65.78$ & \kms \\
    $V_{ej, GC}$ & $698^{+35}_{-27}$  & \kms \\ % ejection  velocity of trajectories that cross within 100pc of SagA*
    $V_{crossing}$ & $486^{+100}_{-151}$ & \kms %plane crossing velocity
    \end{tabular}
\end{table}

\subsubsection{Photometry}\label{subsec:phot_fit}
In this Section, we revise the photometric properties of \candidate by
analyzing the photometric measurements from a variety of surveys with
large wavelength coverage, and fitting the observations with isochrone models. 

Given the nature of this source as a relatively bright object, it has been
 covered by several different surveys. In this Section, we gather photometric 
observations from \gaia, SDSS, SkyMapper, and 2MASS, a summary of which 
is presented in Table \ref{tab:photometric_measurements}. %There is also coverage from WISE, however, we decided not to include those bands as the W3 photometry is likely a spurious detection, being abnormally bright with respect to other bands. 

%To model the photometry, we use the \software{minimint} software to interpolate MIST isochrones. The data we model are the observed magnitudes $m_i$ where $i$ corresponds to the $i$-th band. The iochrones provide us with magnitudes, surface gravities and effective temperatures as a function of stellar age, mass, metallicity, and band-pass $M(\log10(\mathrm{age}), m, [\mathrm{Fe/H}], i)$. We assumed uncorrelated Gaussian uncertainties for the observed magnitudes, which makes our model:
We model the photometry by interpolating MIST isochrones \citep{mist1, mist2, mist3, mist4, mist5} with the 
\software{minimint} package \citep{minimint}. Our dataset consists of the observed 
magnitudes $m_i$, where the index $i$ labels each photometric band. 
The isochrones provide predicted magnitudes, surface gravities, 
and effective temperatures as functions of stellar age, mass 
(${\mathcal M}$), metallicity, and band-pass. Assuming uncorrelated 
Gaussian uncertainties on the observed magnitudes, our model then becomes:

\begin{equation}
\begin{split}
m_i  & \sim  {\mathcal N}\bigg( M (\mathrm{age},{\mathcal M},
[\mathrm{Fe}/\mathrm{H}], i) + \\
 &  + 5 \log_{10} D_{\mathrm{hel}} -5 + k_i E(B-V), \sqrt{\sigma_i^2 + 
 \sigma_{\mathrm{sys}}^2}\bigg),
\end{split}
\label{eq:phot_model}
\end{equation}

%Here, $\sigma_i$ denotes the uncertainty on the magnitude measurement in band $i$, $\sigma_{\mathrm{sys}}$ is an additional systematic scatter about the model, $D_{\mathrm{hel}}$ is the heliocentric distance to the star, and $k_i$ is the extinction coefficient in filter $i$\footnote{Taken from \url{[http://www.mso.anu.edu.au/\~brad/filters.html}}]. Beyond the purely photometric model in Eq.~~\ref{eq:phot_model}, we also adopt a second spectro-photometric model, which augments Eq.~~\ref{eq:phot_model} by incorporating Gaussian constraints on $\log g$, and $\teff$ from the spectroscopic measurement provided by DESI, i.e.\ by multiplying the likelihood by normal‐distribution terms for each of these parameters.

%We adopt the following priors for the parameters: uniform distribution on (linear) $\mathrm{age} \sim {\mathcal U}(10^5, 1.2\times10^{10})$, Salpeter IMF prior for the stellar mass from ${\mathcal M}=0.1$\,M$_\odot$ to ${\mathcal M}=5$\,M$_\odot$, distance is given by the parallax, which is constrained by a Gaussian prior around the \gaia measurement, and a Gaussian prior is given for $[\mathrm{Fe}/\mathrm{H}]$ around the corrected value from DESI, for extinction we adopt a Gaussian prior around the \citet{sfd_og} value with the recalibration given in \citet{sfd_og}. The posterior of the model is sampled using the nested sampling implementation in \software{dynesty} \citep{dynesty, dynesty_sergey}.
Here, $\sigma_i$ denotes the uncertainty in the magnitude measurement 
in band $i$, $\sigma_{\mathrm{sys}}$ represents an additional 
systematic scatter about the model, $D_{\mathrm{hel}}$ is the heliocentric 
distance to the star and $k_i$ is the extinction coefficient in filter 
$i$\footnote{Taken from \url{[http://www.mso.anu.edu.au/\~brad/filters.html}}.
In addition to the purely photometric model in Eq.~~\ref{eq:phot_model}, 
we also adopt a spectro-photometric model that, on top of 
Eq.~~\ref{eq:phot_model}, incorporates Gaussian constraints on $\log g$ 
and $\teff$ from the DESI spectroscopic measurements, i.e., by multiplying 
the likelihood by normal–distribution terms for each of these parameters.

\begin{figure*}
    \centering
    \includegraphics[width=\linewidth]{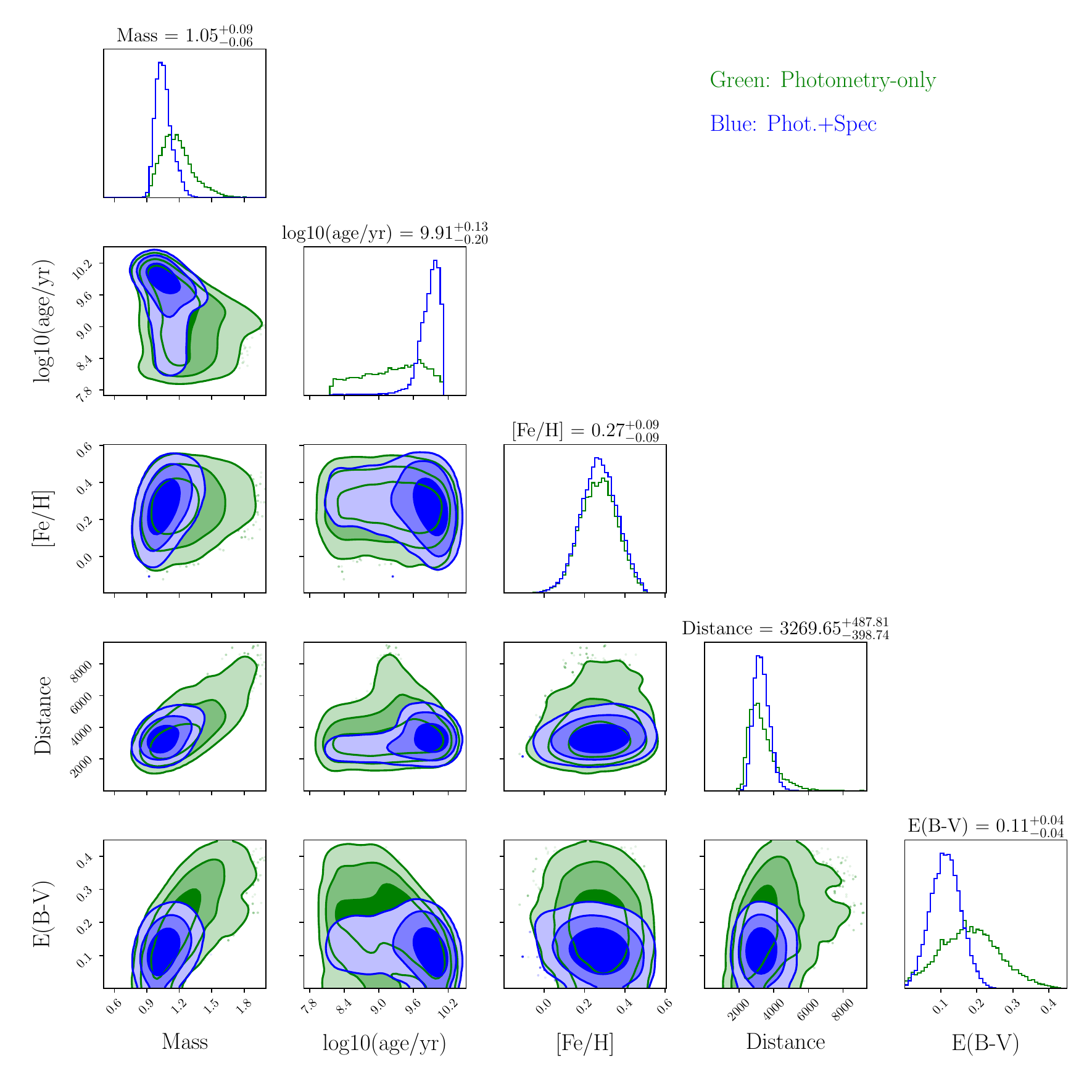}
    \caption{Corner plot of the posterior distributions for stellar 
    mass $M_\star$, $\log_{10}(\mathrm{Age/yr})$, $\mathrm{[Fe/H]}$, 
    distance $d$, and reddening $E(B{-}V)$. Green contours are from 
    the photometry-only fit; blue contours include photometry and the 
    DESI \software{rvspecfit} measurements of \teff\ and $\log g$. Contours 
    indicate the 68\%, 95\%, and 99.7\% credible regions; diagonal panels 
    show the corresponding 1D marginalized posteriors. Titles show the median 
    and 16th and 84th percentiles for the spectro-photometric fit. }

    \label{fig:posterior_stellar}
\end{figure*}

We adopt the following priors for our parameters: a uniform prior on 
(linear) stellar age, where, assuming no star formation in the halo, the lower bound is taken to be the median flight time to the disk, while the upper bound is the Hubble time, i.e., $\mathrm{age}\sim\mathcal{U}(1.5\times10^{8},\,1.2\times10^{10})$; 
a Salpeter-IMF prior for the stellar mass over 
${\mathcal M}=0.1\,M_\odot$–$5\,M_\odot$; a Gaussian prior on distance based 
on the \textit{Gaia} parallax; a Gaussian prior on $[\mathrm{Fe}/\mathrm{H}]$ 
centred on the DESI‐corrected value; a uniform prior in extinction from 0 
to 5 times the reported extinction in \citet{sfd_og}, using the recalibration 
given in \citet{sfd_recal}. The posterior distribution is sampled with the 
nested-sampling implementation in \software{dynesty} 
\citep{dynesty,dynesty_sergey}.   The implementation of this fitting procedure is publicly available here \footnote{https://github.com/mncavieres/mistfit}, and can be used to fit MIST isochrones using any available bands supported by minimint along with the atmospheric parameters, when available. The posterior samples are shown in Figure \ref{fig:posterior_stellar}, while the best fit parameters are reported in Table~\ref{tab:nested_fit}. 

In Figure \ref{fig:posterior_stellar}, the photometric model is shown in green, and we observe a degeneracy between the stellar mass and the distance, as is typical for photometric data.  We derive a stellar mass of $1.17^{+0.38}_{-0.21}M_{\odot}$ and \feh$= 0.27^{+0.17}_{-0.18}$. The age is weakly constrained, 
with the posterior spanning from 170\,Myr to 11\,Gyr (16th to 84th percentiles), stretching as far as the prior allows. The estimated distance is $3.2^{+2.2}_{-0.9}$\,\kpc, in agreement with the \gaia\ parallax-based distance of $3.2^{+1.1}_{-0.6}$\,\kpc. These results are consistent with \candidate\ being a main-sequence star, on the full range of ages between the zero-age and terminal-age main sequence for a  $1.17^{+0.38}_{-0.21}M_{\odot}$ star remains compatible with the data, meaning up to $11$\,Gyr. 

\begin{figure}
    \centering
    \includegraphics[width=\linewidth]{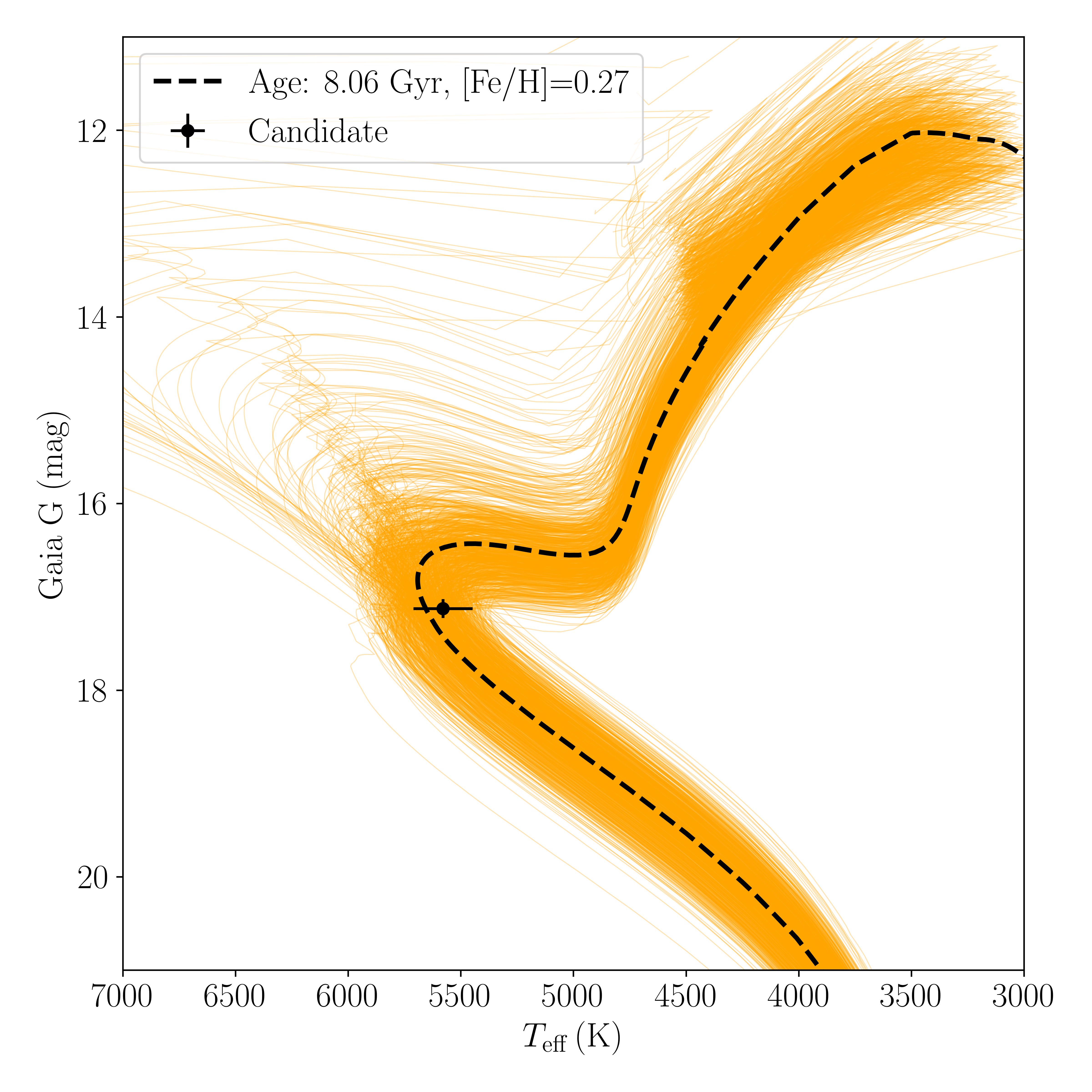}
    \caption{Hertzsprung–Russell diagram of \candidate. The orange curves show 1000 samples from the posterior of the spectro-photometric model (see Figure \ref{fig:posterior_stellar}). Best fit isochrone is shown as a dashed line.}

    \label{fig:hrd}
\end{figure}

For the spectro-photometric model, we find overall agreement with the photometric model, but with improved constraints, particularly on mass and age. While \feh\ remains similarly constrained. This model has a significantly improved age constraint, placing it between $5.12$ - $10.96$\,Gyr corresponding to the 16th and 84th percentiles. We find that \candidate is an old main-sequence star, as evidenced by the star's position in the Hertzsprung-Russell diagram shown in Figure \ref{fig:hrd}, and with 98\% of the posterior samples being consistent main-sequence solutions, while the other 2\% corresponds to it being an evolved sub-giant branch star. While the star is consistent with both, a larger part of the posterior aligns with a main-sequence nature.

We note that if we extend the prior in age towards younger ages, the posterior on the spectro-photometric model also shows a solution for a 14 Myr old star; we do not include this age range as it would imply that the star was formed in the MW halo.

\begin{table}[ht]
  \centering
  \caption{Photometric measurements}
  \label{tab:photometric_measurements}
  \begin{tabular}{lc}
    \hline
    Band            & Magnitude \\
    \hline
    \gaia G  & $17.12 \pm 0.002$ \\
    \gaia BP  & $17.57 \pm 0.011$ \\
    \gaia RP  & $16.53 \pm 0.006$ \\
    SDSS u         & $19.53 \pm 0.036$ \\
    SDSS g         & $17.73 \pm 0.005$ \\
    SDSS r         & $17.08 \pm 0.005$ \\
    SDSS i         & $16.86 \pm 0.005$ \\
    SDSS z         & $16.77 \pm 0.013$ \\
    % W1        & $15.31 \pm 0.041$ \\
    % W2        & $15.38 \pm 0.126$ \\
    % W3        & $11.66 \pm 0.197$ \\
    % W4        & $8.181 \pm 0.197$ \\
    % SkyMapper u    & $19.34 \pm 0.053$ \\ in the end I didn't use these bands because they are the same as SDSS, so why would I do it. 
    % SkyMapper g    & $17.53 \pm 0.007$ \\
    % SkyMapper r    & $17.10 \pm 0.006$ \\
    % SkyMapper i    & $16.86 \pm 0.006$ \\
    % SkyMapper z    & $16.78 \pm 0.008$ \\
    2MASS J        & $15.95 \pm 0.107$ \\
    2MASS H        & $15.37 \pm 0.127$ \\
    2MASS Ks     & $15.40 \pm 0.197$ \\
    $E(B-V)_{sfd}$        & $0.08590$ \\
    \hline
  \end{tabular}
\end{table}

\begin{table}[ht]
\centering
    \caption{Parameters obtained from fitting MIST isochrones to DESI HVS1 photometry and by combining with spectroscopic constraints. Values reported correspond to the median with 16th and 84th percentiles for the lower and upper limits.}
    \label{tab:nested_fit}
\begin{tabular}{l c c l}
\hline

Parameter & Value & Value & Unit \\
 & Photometric & Spectro-Photometric & \\
\hline
\\
\vspace{3pt} 
Mass & $1.17^{+0.38}_{-0.21}$ & $1.05^{+0.09}_{-0.06}$ & $M_{\odot}$ \\
\vspace{3pt}
$\log_{10}$(age/yr) & $9.29^{+0.75}_{-1.06}$ & $9.91^{+0.13}_{-0.20}$ & dex \\
\vspace{3pt} 
\feh & $0.27^{+0.17}_{-0.18}$ & $0.27^{+0.09}_{-0.09}$ & dex \\
\vspace{3pt} 
Distance & $3.18^{+2.21}_{-0.86}$ & $3.27^{+0.49}_{-0.40}$ & kpc \\
\vspace{3pt} 
E(B-V) & $0.18^{+0.17}_{-0.15}$ & $0.11^{+0.04}_{-0.04}$ & mag \\

\hline
\end{tabular}
\end{table}

\subsubsection{Kinematics}\label{subsec:kin}

The kinematic properties of \candidate\ are summarized in Table~\ref{tab:desi_hvs1}. The star is found at $2.15 \pm 0.37$ \kpc below the plane of the disk, and at a distance of $6.43 \pm 0.07$ \kpc from the GC moving at $314 \pm 65$ \kms. The star is currently falling back to the GC, with a flight time since the previous plane crossing of $211 \pm 93$ Myr. 

\begin{figure}
    \centering
    \includegraphics[width=\linewidth]{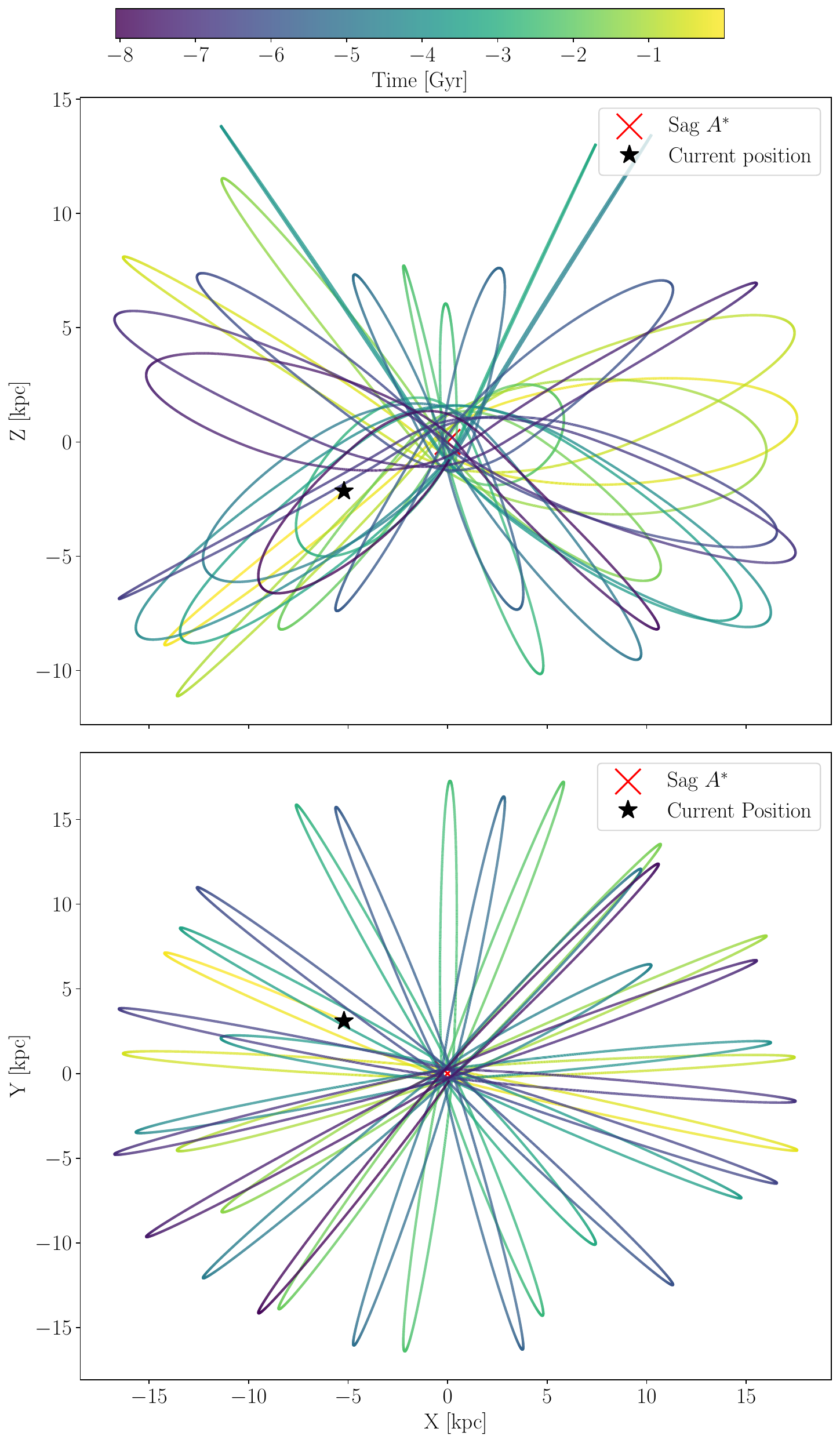}
    \caption{Example orbit of \candidate integrated over the median lifetime  in the XZ (top) and XY plane (bottom), in the Galactic Cartesian coordinate system, colured by the relative time with 0 being the current observed position of the star. The orbit is backward integrated for 8.05 Gyr, given by the median on the age posterior for the spectro-photometric model described in \ref{subsec:phot_fit}.}
    \label{fig:orbit_candidate}
\end{figure}

Given the constraints from the spectro-photometric model described in Section \ref{subsec:phot_fit} and shown in Table \ref{tab:nested_fit}, we integrate the orbit of \candidate\ backwards over the median stellar lifetime of 8.05 \, Gyr, using the \agama\ library. The resulting orbit is shown in Figure~\ref{fig:orbit_candidate}. The top panel shows the orbit in the XZ plane. It reveals chaotic motion with many disk crossings, as expected for ejections via the Hills mechanism \citep{2025MNRAS.542..322P}. The bottom panel shows the orbit in the XY plane and highlights its high eccentricity and repeated close pericenter passages. 

We note that the plotted orbit is only one of many consistent with the observational uncertainties of \candidate. A further caveat is that the adopted Galactic potential is rigid and non-evolving. In reality, the MW has grown in mass through mergers \citep{2025ApJ...991...36Z} and is currently being perturbed by the LMC system \citep{gc2019, 2021Natur.592..534C, 2023Galax..11...59V, yo_lmc}. In addition, the bar can have strong effects on orbits, particularly within resonances \citep{2025MNRAS.542.1331D}. Tests in an evolving barred potential from \citet{hunter_pot} show no significant effect on the orbit of \candidate. Nevertheless, the orbit shown in Figure~\ref{fig:orbit_candidate} should be regarded as an approximate reference, since it neglects both the secular mass growth of the MW and the ongoing LMC perturbation, as well as the observational uncertainties.

\begin{figure}
    \centering
    \includegraphics[width=\linewidth]{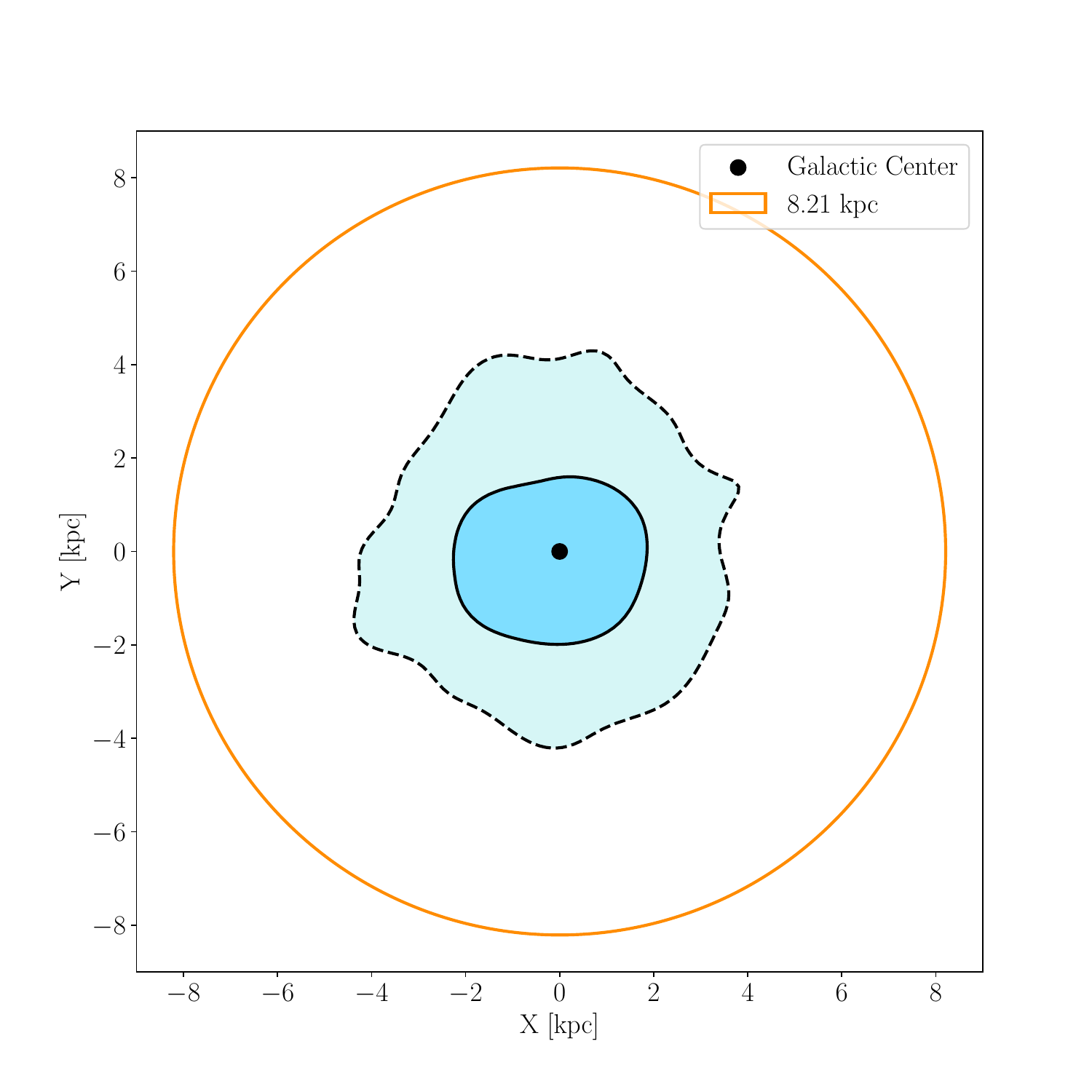}
    \caption{Galactic plane crossing locations for \candidate, considering all crossings that can take place within the lifetime of the star. The 1$\sigma$ and 2$\sigma$ corresponding to regions containing the 68th and 95th percentiles of plane crossings are shown with a continuous and dashed line, respectively. The solar circle is shown in orange for reference.}
    \label{fig:contour_all_crossings}
\end{figure}

The same behavior can be illustrated in the distribution of the plane crossing positions, shown in Figure~\ref{fig:contour_all_crossings}, where the 1$\sigma$ and 2$\sigma$ regions are shown for the Monte Carlo sampling of the orbit, for 100,000 samples and considering all crossings that can take place within the lifetime of the star at the pericenter passage. The posterior peaks at the GC, this is however, less constrained than the distribution of plane crossing locations considering the first plane crossing only. 

The chaotic dynamics evident as well in the positive Lyapunov exponent for around half of our samples, make the exact place of origin difficult to constrain. However, in all allowed orbital realizations the star can be traced back to the Galactic center, owing to its low angular momentum in the Z direction, as shown in Figure~\ref{fig:contour_all_crossings}

\begin{figure}
    \centering
    \includegraphics[width=\linewidth]{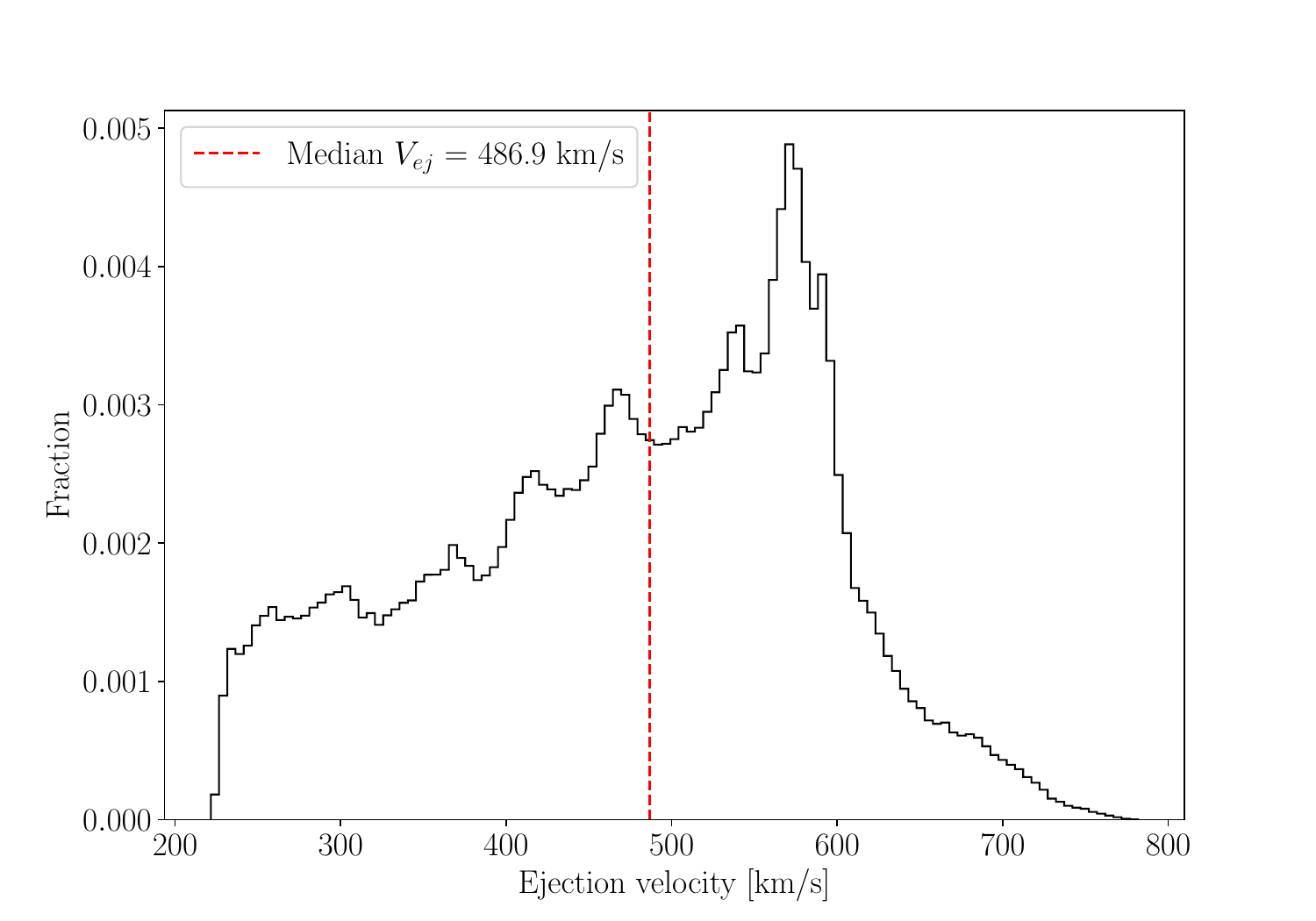}
    \caption{Ejection velocity spectrum from 100.000 orbit realizations integrated back by the median lifetime of the star, compensated by the MW rotation curve as given by the \texttt{MilkyWayPotential2022}.  }
    \label{fig:ej_vel_spectrum}
\end{figure}

We estimate the ejection velocity at a given disk location by subtracting the local circular rotation velocity from the total velocity at that XY position. To do so, we adopt the rotation curve evaluated at each XY position from \texttt{MilkyWayPotential2022}, and compute the ejection-velocity spectrum over 100,000 realizations of backward integration of \candidate, while sampling over the observational uncertainties in proper motion, radial velocity, and distance, and include all crossings allowed within the median lifetime of the star. The resulting ejection velocity spectrum is shown in Figure \ref{fig:ej_vel_spectrum}, with a median ejection velocity of $V_{ej} = 486^{+100}_{-151}$\kms.

\begin{figure}
    \centering
    \includegraphics[width=\linewidth]{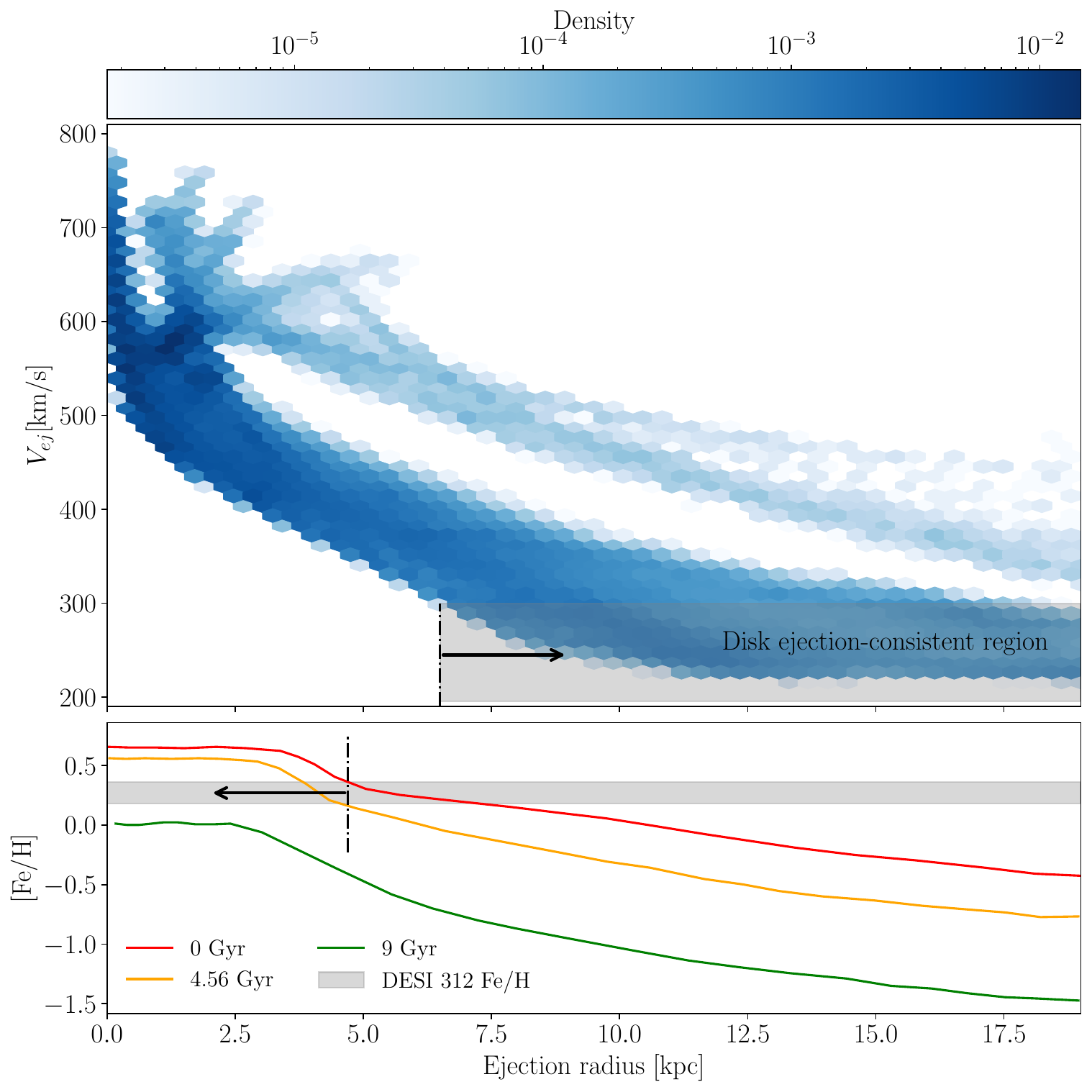}
    \caption{Upper panel: Ejection velocity from the disk versus Galactocentric distance of the ejection point, consistent with the kinematics of \candidate. Velocities and distances are computed by sampling observational uncertainties over 100,000 orbit realizations, integrated backward over the star’s median lifetime; points are colored by the normalized density of orbit crossings in the radius–velocity plane. Lower panel: Fe abundance profiles of the gas 9 Gyr ago (green), 4.56 Gyr ago (yellow), and at present (red), from \citep{2023MNRAS.523.2126P} models. }

    \label{fig:vej_vs_radii}
\end{figure}

The minimum ejection velocity allowed is 221 \kms, which corresponds to an ejection taking place at apocenter with the ejection direction and velocity tuned to eliminate the disk rotation and place the star in the low angular momentum orbit that is currently observed.  This can be observed in the upper panel of Figure \ref{fig:vej_vs_radii}, which shows the ejection velocity as a function of radii, where the lowest ejection velocities (221 \kms) all correspond to apocenter passages beyond 10kpc. On the other side, most of the plane crossings take place at low radii, with the highest density of crossings observed below 2 kpc (also shown in Figure \ref{fig:contour_all_crossings}), here the velocities exceed 500 \kms and reach up to 800 \kms. 

The substructure visible in the upper panel of Figure~\ref{fig:vej_vs_radii} above $600$\kms at radii $\sim 2$\kpc is primarily a result of the uncertainty in the spectroscopic distance from \textit{Specdis}, which maps directly onto uncertainties in both the orbital energy and the vertical component of the angular momentum. For solutions above the locus at ($600$\kms, 2\kpc) the inferred phase-space coordinates permit higher-$L_z$, hence more regular orbits. In contrast, solutions below this locus are restricted to low-angular-momentum, high-energy trajectories that are fully chaotic. Consequently, Hills ejecta consistent with the observed parameters are generically placed on plunging, chaotic orbits \citep[eg.,][]{2025MNRAS.542..322P}.

\begin{figure*}
    \centering
    \includegraphics[width=\linewidth]{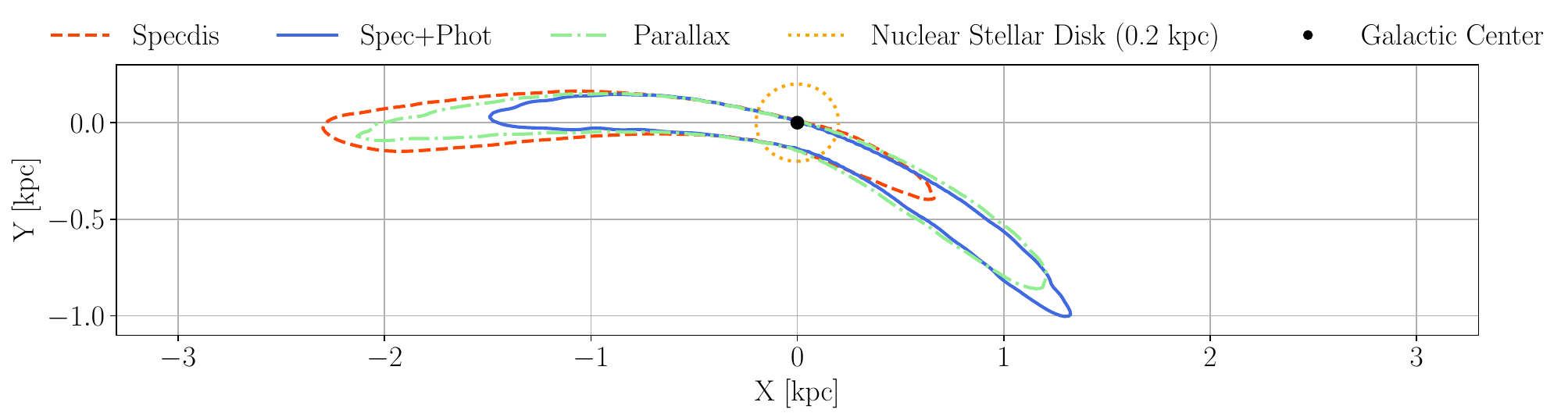}
    \caption{Comparison of Galactic plane crossings for the different distance estimates available for \candidate. For each estimate, the $2\sigma$ region is shown, obtained using the method described in Section \ref{sec:methods}, with the sampling increased to 100.000 Monte Carlo realizations. \candidate shows consistency with GC origin for the first crossing with all available distance estimations. }
    
    \label{fig:plane_crossing_comp}
\end{figure*}

The previous analysis was done using the distances from \textit{SpecDis} catalogue, however \candidate has a well constrained parallax of $0.261 \pm 0.082$ mas, which translates into a distance of $3.82_{-0.91}^{+1.75}$ \kpc, consistent at the 1$\sigma$ level with the spectroscopic distance ($D_{hel} = 4.78 \pm 0.83$ \kpc) reported on Specdist. Recomputing the posterior distribution on the first plane crossing position with all 3 distance estimations, namely: the spectro-photometric model shown in Figure \ref{fig:posterior_stellar}, the \textit{Specdis} neural network distance and the parallax distance; we find that the candidate is consistent with a GC origin for the first crossing for any of the distances used as shown in Figure \ref{fig:plane_crossing_comp}.

%Alternatively, if we assume that the ejection took place at 3\pc from Sgr~A$^\ast$ the ejection velocity reaches an estimated $698^{+35}_{-27}$ \kms. 

\section{Discussions} \label{sec:discussions}

In this Section we discuss the alternatives to the Hills mechanism to explain the properties of \candidate. The young massive cluster ejection is discussed in Subsection \ref{subsec:ymc}. Dynamical disk ejections are discussed in Subsection \ref{subsec:disk_ej}, origin in the LMC and Sagitarius Dwarf are discussed in Subsections \ref{subsec:lmc_origin} and \ref{subsec:sgr_dwarf}.

\subsection{YMC}\label{subsec:ymc}

Among the possible mechanisms capable of accelerating a star to the $486.9$ \kms ejection velocity of \candidate, dynamical interactions in young massive clusters (YMCs) are a leading candidate (see \citet{ymc_review} for a review). Such interactions, along with core-collapse supernovae \citep[e.g.,][]{1961BAN....15..265B, 1998A&A...330.1047T, 2025OJAp....8E..85W}, are typically invoked to explain the origin of main-sequence “runaway stars.” These objects are conventionally defined as having ejection velocities exceeding $40$ \kms \citep{1961BAN....15..265B}—an arbitrary threshold chosen to distinguish them from “walkaway stars,” whose velocities just exceed the escape velocity of their natal environment \citep{ymc_review}. However, both mechanisms struggle to produce ejection velocities above $\sim$200 \kms \citep{2000ApJ...544..437P, 2012ApJ...751..133P, 2016A&A...590A.107O, fraser_runaways, 2025arXiv250614273E} with ejections above 200 \kms being heavily outnumbered by ejections on the order of $\sim 10$ \kms \citep{2000ApJ...544..437P, 2011MNRAS.414.3501E, 2025OJAp....8E..85W, 2012ApJ...751..133P}. 

There are two well known YMCs in the galactic center that lie within the posterior for the first place crossing shown in Figure \ref{fig:plane_crossing_comp}, the Arches \citep{1995AJ....109.1676N} and Quintuplet cluster \citep{1983PASJ...35..101K}. The age of the Arches cluster is found to be between $2.0-3.3 $ Myr, given the observed main sequence turnoff, the population of supergiant stars, and the lack of H-free Wolf-Rayet stars \citep{2018A&A...617A..65C}, in a similar way, the age of the Quintuplet cluster has this age estimated to be $3.0 - 3.6$Myr \citep{2012A&A...540A..14L}, both are inconsistent with the estimated age for \candidate at  $\sim 5.12 - 10.9$ Gyr, thus we reject both the Arches and Quintuplet clusters as a possible progenitor for \candidate, since they are both unlikely to generate the required ejection velocity and have an inconsistent age.

The ejection velocity spectrum of \candidate is shown in Figure \ref{fig:ej_vel_spectrum} from which a median ejection velocity of 486 \kms is found, greatly exceeding typical ejection velocities from clusters \citep{2012ApJ...751..133P}, making the cluster ejection hypothesis highly unlikely for \candidate, in line with findings that dynamical and supernovae ejections cannot explain the HVS population \citet{2012ApJ...751..133P, fraser_runaways}. Note that this velocity is higher than the galactic reference frame velocity ejection since the later should also cancel out the MW rotation in order to match the angular momentum of \candidate.

\subsection{Disk ejections}\label{subsec:disk_ej}

Previous studies have identified a large number of fast objects that cannot be traced back to the GC and are instead associated with an origin in the MW disk, such as some of the candidates presented in  \citet{hvs_survey, kreuzer_2020}. To assess the possibility of \candidate from being a disk ejection, we simulate a population of MW disk ejecta. 

We generate birth sites from a geometrically thin, axisymmetric exponential disk and endow each star with the local circular motion of a realistic Milky Way potential before perturbing it with small dispersions and an isotropic kick. Cylindrical galactocentric radii are drawn from the surface–density profile ($\Sigma(R)\propto e^{-R/R_d}$), i.e. from the area–corrected radial PDF ($p(R)\propto R,e^{-R/R_d}$), truncated at ($R_{\max}$); azimuths are uniform in ([0,2$\pi$)) \citep{2008ApJ...673..864J}. Vertical positions are sampled from a symmetric Laplace distribution about the midplane, ($p(z)\propto \exp(-|z|/h_z)$), appropriate for a thin disk. At each sampled ((x,y,z)) we evaluate the circular speed from the adopted Milky Way potential, and assign initial velocities around the circular orbit. These are converted to Cartesian components and we include a small additional in–plane Gaussian turbulence ($\varepsilon_{x,y}\sim\mathcal{N}(0,\sigma_{\rm turb})$) to mimic non–circular motions. Unless stated otherwise, we adopt fiducial values ($R_d=2.6,\mathrm{kpc}$), ($R_{\max}=15,\mathrm{kpc}$), ($h_z=0.1,\mathrm{kpc}$), ($\sigma_{v_R}=\sigma_{v_z}=10$\kms), and ($\sigma_{\rm turb}=15,\mathrm{km,s^{-1}}$) \citep{2017MNRAS.465...76M, 2019ApJ...871..120E}. The gravitational field and hence ($v_c(R)$) are taken self–consistently from the \texttt{MilkyWayPotential2022} within AGAMA, ensuring that the rotation curve reflects the same mass model used to integrate trajectories.

To model cluster–ejection mechanisms (few–body interactions and supernova kicks), we then impart to each star an additional isotropic velocity kick drawn uniformly on the sphere and speeds sampled from a broken power–law distribution, $p(v_{\rm k})\propto v_{\rm k}^{-8/3}$ between (0.1 - 300 \kms) and $p(v_{\rm k})\propto v_{\rm k}^{-3/2}$, following the dynamical ejection velocity distribution from \citet{2012ApJ...751..133P}. This procedure yields a population of disk ejecta whose birth radii, vertical distribution, and pre–ejection kinematics reflect a thin exponential disk embedded in a realistic Galactic potential, while the kick spectrum and isotropy encode the cluster-driven ejection channel with an upper limit in ejection velocity of 300 \kms. The resulting initial conditions are forward–integrated 1\,Gyr to predict present–day positions and velocities for comparison with the observed high–speed candidates.

From the simulated dynamical disk ejections, we find that it is possible to produce stars with orbits similar to that of \candidate. However, placing a star in such an orbit within the halo is relatively unlikely, as it requires fine-tuning of both the ejection velocity and direction to achieve a radial trajectory consistent with that observed for \candidate. In our simulation, approximately 2\% of disk ejections result in $|z| > 2$\kpc. Furthermore, only about $10^{-5}$ of all ejections lead to a configuration compatible with the 16th and 84th percentiles of \candidate's energy and angular momentum. This low fraction reflects the stringent requirement that a star must be ejected with both the appropriate velocity and direction to lose angular momentum while retaining relatively high energy.

%Considering the upper limit of 20 hyper-runaway O-type stars predicted by \citet{2012ApJ...751..133P}, extending this estimate to all stellar types yields a total of approximately 100 stars, of which about $10^{-3}$ would be expected to occupy an orbit compatible with \candidate. This estimate corresponds to an upper limit as it does not account for stellar lifetimes, and neglects the fact that the fastest ejections predominantly occur for high-mass O- and B-type stars (see Figure 3 in \citet{2012ApJ...751..133P}), while \candidate is a solar-type star.
We can estimate an upper limit on the number of runaway stars that may lie in orbits compatible with \candidate by first assuming that all star formation in the galaxy has taken place in clusters, with a median star formation rate of $\Psi = 2 M_\odot yr^{-1}$ \citep{2022ApJ...941..162E}, and a relatively high cluster mass capable of ejecting stars at high velocity $M_{cl} = 10^4 M_\odot$, then within the time $T = 8$ Gyr, which corresponds to the lookback time given by the 16th and 84th percentiles in age of \candidate, we will have a total number of clusters formed:

\begin{equation}
    N_{cl} = \frac{\Psi T}{M_{cl}}
\end{equation}

Given \citet{2012ApJ...751..133P} most realistic model, we can expect 0.02 stars/cluster to be ejected with $v_{ej} > 300$ \kms. Additionally, we need the star to survive to the present day, which requires the mass to be $m \leq 1M_\odot$ given that \candidate is about to exit the main sequence with a solar mass, therefore considering Figure 4 in \citet{2012ApJ...751..133P}, we can expect $\sim 30\%$ of runaways to be within this mass regime, as a result, the rate of runaways becomes $\epsilon = 0.06$ stars/cluster, of which as previously described the ratio \fcomp will have compatible orbits. Hence, the number of runaways with orbits compatible with \candidate that have accumulated within the lookback time of ages compatible will be:

\begin{equation}
    N_{rw} = N_{cl} \times \epsilon \times f_{comp} = \frac{\Psi T}{M_{cl}}\times \epsilon \times f_{comp} \sim 7
\end{equation}

%Note that this upper limit considers that the totality of star formation occurs in massive clusters with a mass of $10^4 M_{\odot}$, while in reality the cluster population is given by the cluster mass function, which peaks at a mass of $10^5 M_{\odot}$ \citep{2008ASPC..388..279L, 2009MNRAS.394.2113G, 2023A&A...672A.187J}. As our assumption is that the number of runaways depends on the number of clusters, therefore, increasing the cluster mass would, in principle, decrease the number of runaways by a factor of 10, if $\epsilon$ stays constant. Studying the variation of $\epsilon$ as a function of cluster mass is outside the scope of this work. 

Note that this upper limit assumes that all star formation occurs in massive clusters with a mass of $10^4M_{\odot}$, whereas in reality the cluster population follows a cluster mass function that peaks at $10^5M_{\odot}$ \citep{2008ASPC..388..279L, 2009MNRAS.394.2113G, 2023A&A...672A.187J}. Under our assumption that the number of runaways scales with the number of clusters, increasing the characteristic cluster mass by a factor of 10 would, for fixed total stellar mass and constant $\epsilon$, reduce the predicted number of runaways by the same factor. A detailed study of the dependence of $\epsilon$ on cluster mass is beyond the scope of this work, but it would be a significant improvement on our estimation as this parameter is expected to be an increasing function of mass. 

Additionally, this estimate does not account for detectability by Gaia or DESI, and it includes all stars with masses lower than \candidate, which are less likely to be detected by either survey. Thus, it should be regarded as an upper limit on the number of stars that could be identified via this mechanism.

Moreover, for the star to be ejected at a velocity consistent with runaway ejection processes ($V_{ej}<300$ \kms in the most extreme cases), it would need to originate in the outer disk. As shown in the upper panel of Figure \ref{fig:vej_vs_radii}, ejection velocities below 300 \kms are observed only beyond 10 \kpc, within the shaded region where the most extreme runaways would lie. Furthermore, considering the Fe abundance profile in the lower panel of Figure \ref{fig:vej_vs_radii}, the formation of stars with metallicities consistent with our candidate (\feh $\sim 0.3$) beyond 4 \kpc is not expected from chemical-evolution models of the galaxy, even allowing for the lower age limit of \candidate at $\sim 5$ Gyr; this is especially unlikely given the median age of \candidate at $\sim 8$ Gyr. We therefore consider the region within 4 \kpc to be consistent with the \candidate \feh. Consequently, within the region ($R\lesssim 4$ \kpc) where the metallicity is consistent with \candidate, the required ejection velocity would have to exceed the runaway threshold ($V_{ej}\ge 300$ \kms), thereby almost completely ruling out runaway mechanisms as a viable origin.

%\zp{up to you, but given that we can explain the substructure in the $R,\ v_{ej}$ panel of figure 8 i would include this (that there's a parallax dependance which translates into a lz and energy dependance, and allows the higher orbits in the plot access regular orbits, whilst below that they're all chaotic) - i think this is a crucial piece of understanding for these hills ejecta, that they should all be on chaotic diving orbits (as we showed in Penoyre, Rossi and Stone 2025, just inverting the direction of time)}

% Under th

Therefore, while this remains an extremely unlikely scenario, we do not rule out completely the possibility that \candidate was ejected from an unknown (and possibly now-dissolved) cluster with an outlier metallicity in the disk, with the precise velocity and direction necessary to reach its current radial bound orbit, in particular if ejection velocities from runaway processes can exceed 300 \kms, which to date has not been observed.

%If we assume that the ejection mechanism for disk ejecta is that of cluster ejections (ie., n-body interactions and super novae ejections) we simulate such ejections by generating 

\begin{figure}
    \centering
    \includegraphics[width=\linewidth]{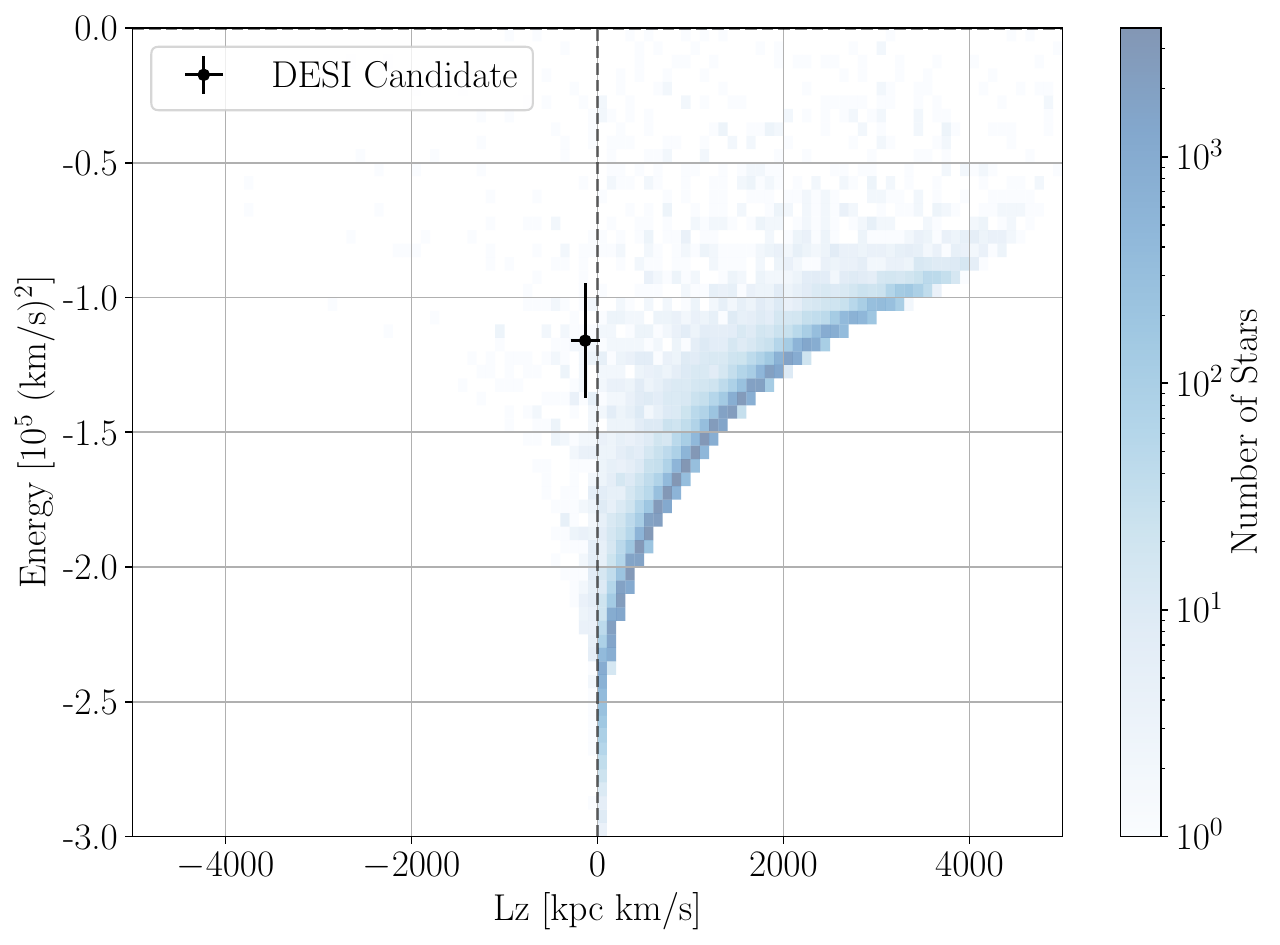}
    \caption{Integral of motion space for 100.000 simulated dynamically ejected runaway stars and hyper-runaway stars following the \citep{2012ApJ...751..133P} velocity distribution. \candidate is shown in black, with energy and angular momentum compatible at the $1\sigma$ level with $L_z = 0$, but mainly a counter-rotating angular momentum. }
    \label{fig:integral_motion_space}
\end{figure}

\subsection{Metal-rich globular clusters}\label{subsec:met_rich_gc}

Within the dynamical ejection scenario, three- and four-body interactions in dense stellar clusters can convert binary binding energy into kinetic energy and eject cluster members at high velocities. However, as previously mentioned, N-body simulations show that ejection velocities above $\sim 200~\mathrm{km,s^{-1}}$ are rare in such dynamical encounters and cannot account for the population of hyper-runaway stars with $V_{\rm ej} \gtrsim 400~\mathrm{km\,s^{-1}}$.

A natural class of dense stellar systems in which dynamical ejections can occur is globular clusters. In the Milky Way, $\sim 170$ globular clusters are currently known \citep[e.g.][]{2021MNRAS.505.5978V}, and their metallicities are predominantly subsolar ($\mathrm{[Fe/H]} < 0$). Supersolar metallicities have been proposed for the bulge globular clusters Liller 1 and NGC 6528 \citep{2009A&A...508..695C, 2017A&A...601A..31L}, although in both cases the quoted uncertainties are comparable to or larger than the inferred metallicity excess. 

Considering that the highest-metallicity clusters are typically found in the bulge region \citep[e.g.][]{2024A&A...687A.214G}. Thus, if the existence of a metal-rich tail among bulge globular clusters is confirmed, dynamical ejection from such a system could represent a plausible progenitor channel for \candidate (\feh $\sim 0.27$), particularly if its ejection velocity lies near the low end of the distribution shown in Fig.~\ref{fig:ej_vel_spectrum}.

\subsection{LMC origin}\label{subsec:lmc_origin}
The LMC is the MW most massive satellite, with a mass estimated around 10\% of the MW mass \citep{2010ApJ...721L..97B, 2016MNRAS.456L..54P, 2017MNRAS.464.3825P, 2019MNRAS.487.2685E, 2020MNRAS.495.2554E, 2021ApJ...923..149S, 2021MNRAS.501.2279V, 2022MNRAS.511.2610C, 2023MNRAS.521.4936K}. As such, it is believed to host a supermassive black hole \citep{2017ApJ...846...14B}, which is also ejecting HVS via the Hills mechanism \citep{lmc_hvs}. Hence, it is worth assessing the LMC as a possible progenitor for HVS candidates.

%To explore the possibility of an LMC origin for \candidate, we integrated 1,000 realizations sampling over uncertainties as in \ref{sec:methods}, within a time-dependent MW+LMC potential. This potential models the Milky Way as a simplified system comprising a spherical bulge, a single exponential disk, and a spherical dark matter halo. The LMC is represented as a $1.5\times10^{11}\,M_{\odot}$ spheroid \citep{2023MNRAS.521.4936K}, consistent with the constraints from \citet{agama}. The mutual MW–LMC orbit is computed self-consistently, including dynamical friction via the standard Chandrasekhar formula. This model reproduces the LMC's orbital trajectory compared to the N-body simulations of \citet{gc2019} over the last 0.5\,Gyr. We found that none of the 1,000 realizations cross within 10 \kpc from the LMC center, we thus reject an LMC origin for \candidate.

To test an LMC origin for \candidate, we integrate 10,000 realizations that sample the observational uncertainties (Section \ref{sec:methods}) in a time-dependent MW+LMC potential. The Milky Way is modeled as a spherical bulge, a single exponential disk, and a spherical dark matter halo. The LMC is represented as a $1.5\times10^{11},M_{\odot}$ spheroid \citep{2023MNRAS.521.4936K}, consistent with the constraints from \citet{agama}. The mutual MW–LMC orbit is computed self-consistently and includes dynamical friction via the standard Chandrasekhar formula. Over the last 1\,Gyr, this model reproduces the LMC orbit from the N-body simulations of \citet{gc2019}. None of the 10,000 realizations pass within 40,\kpc of the LMC center, so we reject an LMC origin for \candidate.

\subsection{Sagittarius Dwarf origin}\label{subsec:sgr_dwarf}

The Sagittarius Dwarf galaxy is a dissolving galaxy being tidally disrupted by the MW \citep{sag_discovery}. It is one of the closest and most massive satellites of the MW, and the most metal rich \citep{mw_sat_chem}. As such, it is chemically consistent with the supersolar metallicity of our candidate (\feh $\sim 0.27$), in particular if it was ejected from the core of Sagittarius.

However, considering the Sagittarius Dwarf trajectory from \citet{sag_orbit} within the last 3 Gyr as fixed, and propagating 100,000 realizations of trajectories for \candidate in the same potential used to construct the Sagittarius Dwarf trajectory from \citet{sag_orbit}, which considers an evolving potential with dynamical friction for the LMC + Sagittarius Dwarf. We find that the median closest approach between both objects is $\sim 12.1 \pm 2.7$ \kpc. From a kinematic standpoint, this makes an origin of \candidate\ in the Sagittarius dwarf unlikely, although not completely ruled out, in contrast to the case of the LMC.

\subsection{Previous work}\label{subsec:previous_work}

Recently, \citet{sill_desi} conducted a search using the same dataset employed in this work for a statistical overdensity of low-angular-momentum ($L_z \sim 0$), high-metallicity ($-0.25 < \feh < 0.5$) stars, effectively probing the same parameter space occupied by \candidate. The detection of such an overdensity could not be statistically confirmed. This non-detection places an upper limit of approximately 51 stars that could still be present in DESI, which is consistent with the identification of a single candidate based on the selection function described in Section \ref{sec:methods}.

\citet{hattori25} performed a 6D search that also included bound HVSs, using \gaia DR3 astrometry, distances from Starhorse \citep{2022A&A...658A..91A}, and metallicities derived from Gaia XP spectra in \citet{2025ApJ...980...90H}. Their analysis identified a single HVS candidate: the red giant WINERED HVS1. This candidate is bound, similar to \candidate, and is distinguished from halo stars on radial orbits by its chemical composition, both from \gaia XP-derived metallicities and from a high-resolution spectroscopic follow-up in which the authors chemically tagged WINERED HVS1 to MW bulge populations. Further association with nuclear star cluster trends could be assessed by comparing its abundance trends to those presented in \citet{nsc_chemistry}, particularly the excess $[Na/Fe]$ found in the nuclear star cluster relative to the nuclear stellar disk and bulge populations. A similar analysis could also be performed for \candidate given future high-resolution spectroscopic follow-up.

\section{Conclusions}\label{sec:conclusions}

%In this work, we searched for HVSs using a Monte Carlo orbit integration scheme that combines \gaia astrometry, distances from \textit{Specdis}, and data from DESI DR1. The method does not discriminate between bound and unbound stars and is therefore sensitive to most HVSs ejected from the GC. This sensitivity is achieved by combining kinematics with the \feh measurements in DESI DR1. 
We performed a six-dimensional search for hypervelocity stars in DESI DR1 \citep{desidr1} by combining DESI radial velocities and metallicities with \gaia\ DR3 astrometry and spectrophotometric distances from \textit{Specdis} to build a 6D catalogue containing about 4 million objects. Our selection admits both bound and unbound ejecta and couples backward Monte Carlo orbit integration in a realistic Milky Way potential with a supersolar metallicity requirement to suppress contamination from halo stars in radial orbits. The method identifies candidates that (i) cross the Galactic plane within 1 \kpc of \sag in a significant fraction of realizations and (ii) are chemically inconsistent with Gaia–Sausage–Enceladus or other halo populations.

%The HVS search methodology described in section \ref{sec:methods} yields a single HVS candidate, to which we refer as \candidate. Here we summarize the findings:
Here we summarize the findings from our HVS search:

   \begin{enumerate}

    \item Our search identifies \candidate\ as a strong candidate GC ejecta. It is a $\sim 1\,M_{\odot}$, high-metallicity star (\feh\ $\simeq 0.27$~dex) on a very eccentric orbit ($e\sim 0.97$) with almost purely radial motion, characterised by a normalised angular momentum $L_z/L_c(E) = -0.018\pm0.06$. Its first Galactic-plane crossing lies within 2\kpc of \sag\ in the vast majority of Monte Carlo realisations (see Figure \ref{fig:plane_crossing_comp}). The present-day Galactocentric speed is $V_{\mathrm{gcr}}\simeq 314\pm 66$~\kms, the median ejection speed is $V_{\mathrm{ej}}\simeq 486.9$~\kms, and the inferred GC (ejected within 1pc from Sag A*) ejection speed is $V_{\mathrm{ej,GC}}\simeq 698^{+35}_{-27}$~\kms, values compatible with ejection via the Hills mechanism.

    %item \zp{rearrange the first bit of this sentence} \candidate old main–sequence nature strongly disfavors an origin in known young massive clusters that currently exist in the galaxy, such as the Arches and Quintuplet clusters, which are found within the posterior on the plane crossing position of this candidate. 
    
    \item From the spectro-photometric modelling (Section \ref{subsec:phot_fit}), \candidate\ is an old main-sequence or slightly evolved subgiant star, with an age of $log_{10}(\textrm{age/yr}) = 9.91^{+0.13}_{-0.21}$ and a mass of $1.05^{+0.09}_{-0.06}\,M_{\odot}$. This age range is incompatible with an origin in known young massive clusters in the Galactic Centre region, such as the Arches and Quintuplet clusters (ages $2$--$4$~Myr), even though these systems lie within the posterior on the first plane-crossing position.

    \item The high metallicity measured for \candidate makes it extremely unlikely for it to be a halo star in a radial orbit, particularly from the known accretion event of Gaia-Sausage Enceladus, as well as an ejection from globular clusters.

    \item \candidate required kick speeds for a disk ejection likewise exceed those expected from dynamical or supernova–driven runaways. This, combined with the particular energy and angular space that \candidate occupies, makes it unlikely to originate from a disk ejection. Moreover, given the MW metallicity gradient, stars with metallicities comparable to our candidate are expected to form predominantly in the inner regions of the MW where the required ejection velocity exceeded the runaway regime ($V_{ej} \geq 350$\kms). Thus, within the region where the metallicity is consistent, the necessary ejection velocities almost completely rule out a runaway origin.

   \end{enumerate}

High resolution follow-up spectroscopy is essential both to rule out unresolved binarity that could bias our measurements, and further improve the association to the GC by chemical tagging. Specifically, testing whether \candidate\ exhibits the elevated $[\mathrm{Na}/\mathrm{Fe}]$ trend identified for NSC stars \citep{nsc_chemistry} when compared to other bulge stars. Additionally, a more precise radial velocity measurement would further tighten the kinematic constraints on its origin.

If the origin of \candidate is confirmed through chemical tagging, high resolution spectroscopy would allow to exploit the full potential of \candidate as a tracer of the Galactic Center’s chemical composition, since unlike previously identified A- and B-type HVSs, \candidate is solar-type, enabling detailed chemical analysis of its atmosphere and providing a window into the chemistry of the central regions without the hindrance of dust and crowding.

The analysis presented here can be applied to other large spectroscopic surveys, such as LAMOST \citep{lamost}, APOGEE \citep{apogee}, WEAVE \citep{2012SPIE.8446E..0PD}, and 4MOST \citep{4most}, and the upcoming Gaia DR4 and DESI DR2, which will increase by a factor of three the number of sources to which this analysis can be applied, and potentially increase the number of HVS detected. 

\begin{acknowledgements}
The authors acknowledge Bianca Sersante, Andres Presa, and Konstantinos Kilmetis for their insightful discussions. 

EMR acknowledges support from European Research Council (ERC) grant number: 101002511/project acronym: VEGA\_P. SK acknowledges support from the Science \& Technology Facilities Council (STFC) grant ST/Y001001/1. 
For the purpose of open access, the author has applied a Creative
Commons Attribution (CC BY) licence to any Author Accepted Manuscript version
arising from this submission.\\

Software: \texttt{NumPy} \citep{harris2020array}, \texttt{SciPy} \citep{2020SciPy-NMeth}, \texttt{Matplotlib} \citep{Hunter_2007}, \texttt{Astropy} \citep{astropy:2013, astropy:2018, astropy:2022}
\end{acknowledgements}


\bibliographystyle{aa}
\bibliography{bib}

\end{document}